\documentclass[preprint,12pt]{elsarticle}

\usepackage{amssymb}
\newtheorem{defi}{Definition}

\journal{Brain Research Bulletin}

\begin{document}

\begin{frontmatter}



\title{Reality as simplicity}


\author{Giulio Ruffini}

\address{Starlab Barcelona, C. de l'Observatori, s.n., 08035 Barcelona, Spain}

\begin{abstract}
The aim of this paper is to study the relevance of simplicity and its formal representation as  Kolmogorov or algorithmic complexity in the cognitive sciences. The discussion is based on two premises:  1) all human experience is generated in the brain, 2) the brain has only access to  information. Taken together, these two premises lead us to conclude that all the elements of what we call `reality'  are derived mental constructs based on information and compression, i.e., algorithmic models derived from the search for simplicity in data.  Naturally, these premises apply to humans in real or virtual environments as well as  robots or other cognitive systems.  Based on this, it is  further hypothesized that there is a hierarchy of processing levels where simplicity and compression play a major role.  As applications, I illustrate first the relevance of compression and simplicity in fundamental neuroscience with an analysis of  the Mismatch Negativity paradigm. Then I discuss the applicability to  Presence research, which studies how to produce real-feeling experiences in mediated interaction, and use   Bayesian modeling  to define in a formal way different aspects  of the illusion of Presence. The idea is put forth  that given alternative models (interpretations) for a given mediated interaction, a brain will select the simplest one it  can construct weighted by prior models.
In the final section the universality of these ideas and applications in  robotics, machine learning, biology and education is discussed. I emphasize that there is a common conceptual thread based on the idea of simplicity, which suggests a common study approach. 
\end{abstract}


\end{frontmatter}

\clearpage


\section{Introduction}

\begin{flushright} {\small \em  
 Reality is merely an illusion, albeit a very persistent one. \\
 Everything should be made as simple as possible, but not simpler. \\
A. Einstein
}
\end{flushright}

In this paper I discuss a general paradigm to study cognition based on ideas from algorithmic information and probability theory. It is a brain-centric,  subjective paradigm. While   following a long tradition in theories about modeling and simplicity, its novelty lies perhaps in its application in the cognitive sciences, both natural and artificial,  providing a common framework that brings together many scientific endeavors, from neuroscience to physics to machine learning.  
In this paradigm, the central concepts are information and brains---or, more generally, cognitive systems.  
As we will show, putting the brain in center stage  will lead us to the idea that brains seek to model  the incoming flux of information---and to simplicity as a driving principle. 

In order to illustrate the general applicability of thinking about simplicity, we can start with its application in the philosophy of science. Loosely speaking, science endeavors in observing phenomena by carefully planned experimental work,  extracting the rules or regularities in observed data, and using them to make predictions which can help us falsify and thus improve them. The goal of science is to develop simplifying models with compressive and predictive power  \cite{Chaitin:2004aa}.        A mystery  that has baffled scientists for centuries is why simple models, which by definition are good compressors, are also good at predicting the future. 

A perhaps more intriguing statement is that science is essentially what brains do: cognition is first and foremost model building.  When a scientist---which we could call a professional thinker---or a cognitive system develops a model of what is observed,  a tool for data compression is produced.  A good model can be used to literally compress the collected information: we can just compute  the difference between data and model outputs (e.g., using Huffman coding). A good model will give produce an easy-to-compress difference file with mostly 0's. 

But is this all a model is, a compression engine? In a general sense, a model is a mathematical construct that is designed to correspond to a prototype, which may be physical , biological, social, psychological or even some other mathematical model \cite{Aris:1978aa}.  As such, the purpose of a model is to 1) predict some aspects of the future, 2) influence further experimentation or observation, 3) foster conceptual progress and understanding, 4) assist the axiomatization of the prototyped situation, and 5) foster mathematics and the art of making mathematical modes \cite[p. 78]{Davis:1981aa}. The first aspect, in particular, is somewhat mysterious. Why can we predict the future at all?

What does this   have to do with the elusive concept we call `reality'?   If `reality' belongs to the realm of cognitive systems, then it must  be firmly anchored on the concept of information, which is all cognitive systems, natural or artificial, have access to.   In other words, reality is an information-based construct of our brains---a model---and one we could certainly call an illusion, too. And since, as we will argue, models are to represent information in simple terms, the notion of reality arises, ultimately, from the search for simplicity. In some sense `simplicity' is equivalent to `reality'.   

Information  has now taken the central role in many human activities, and most clearly in  science. Formally,  it is a well defined concept describing to the number of yes/no questions required to specify a physical (or abstract) state and it  owes its existence to the mathematical concept of the bit, the essence of dichotomy and an invention of Leibniz.
 As our brains deal with information from the external universe, a data stream of huge bandwidth is compressed into manageable concepts such as `mass' or `matter', `person' or `justice'.  How does this happen? In existential terms, the initial conclusion a blank brain would reach from is this:   {\em there is information}. In philosophical terms, materialism is replaced by `informationalism' or `modelism', bringing us closer to Platonic idealism, or reality as mathematics.


While the points discussed so far may be intuitive, we need a rigorous definition of simplicity, and we need to explain why simplicity is important to brains.   Fortunately, a precise definition of simplicity, or rather its antonym, complexity,  exists: it is provided by algorithmic information theory,  further described below. 
	
These ideas follow a rather long thread which originate perhaps with Plato and were then  more clearly articulated  by  Leibniz in his {\em Discourse on Metaphysics} (1686)---as pointed out by Chaitin \cite{Chaitin:2004aa}. The importance of simple explanations is also commonly associated with Occam, who in the 14th century stated  {\em ``Pluralitas non est ponenda sine neccesitate''}, which we may translate  loosely as {\em ``models should be no more complex than sufficient to explain the data''}. This is also known as `Occam's razor'. In more recent times, the idea that explanations should be as simple as possible---but no more---was also exposed and practiced by Einstein, to whom is also attributed the statement that reality is just a persistent illusion\footnote{Indeed, the discovery of the theory of relativity---which profoundly separated our daily experience of space and time from their physical description----and quantum mechanics have eroded significantly the seemingly rock solid concept we had of `reality', bringing us closer to Plato's cave.}.  More recently,  the    relevance of the idea of simplicity in the cognitive sciences and its relation to Kolmogorov Complexity has been pointed out \cite{Chater:2003aa,Ruffini:2007aa}.  The present paper aims to  continue in this direction by shifting further the focus from the search for simplicity in the physical universe to the subjective realm of the human brain and model building in cognitive systems.

In the following sections,  the brain-in-the-universe model is analyzed, and the meaning of computation  in this context briefly discussed. This is followed by a  review of the various definitions for simplicity and  a discussion of  why simplicity is an important concept in the context of  neuroscience and  Presence. 
In the last section of this paper I provide an overview of the relation of these ideas to several other fields---including physics, biology, machine learning,  robotics and education---and suggest that a common research framework may be applicable.

\begin{figure*}[t!]

\vspace{-2.5cm}

\hspace{-.7cm} 
\includegraphics[width=15cm]{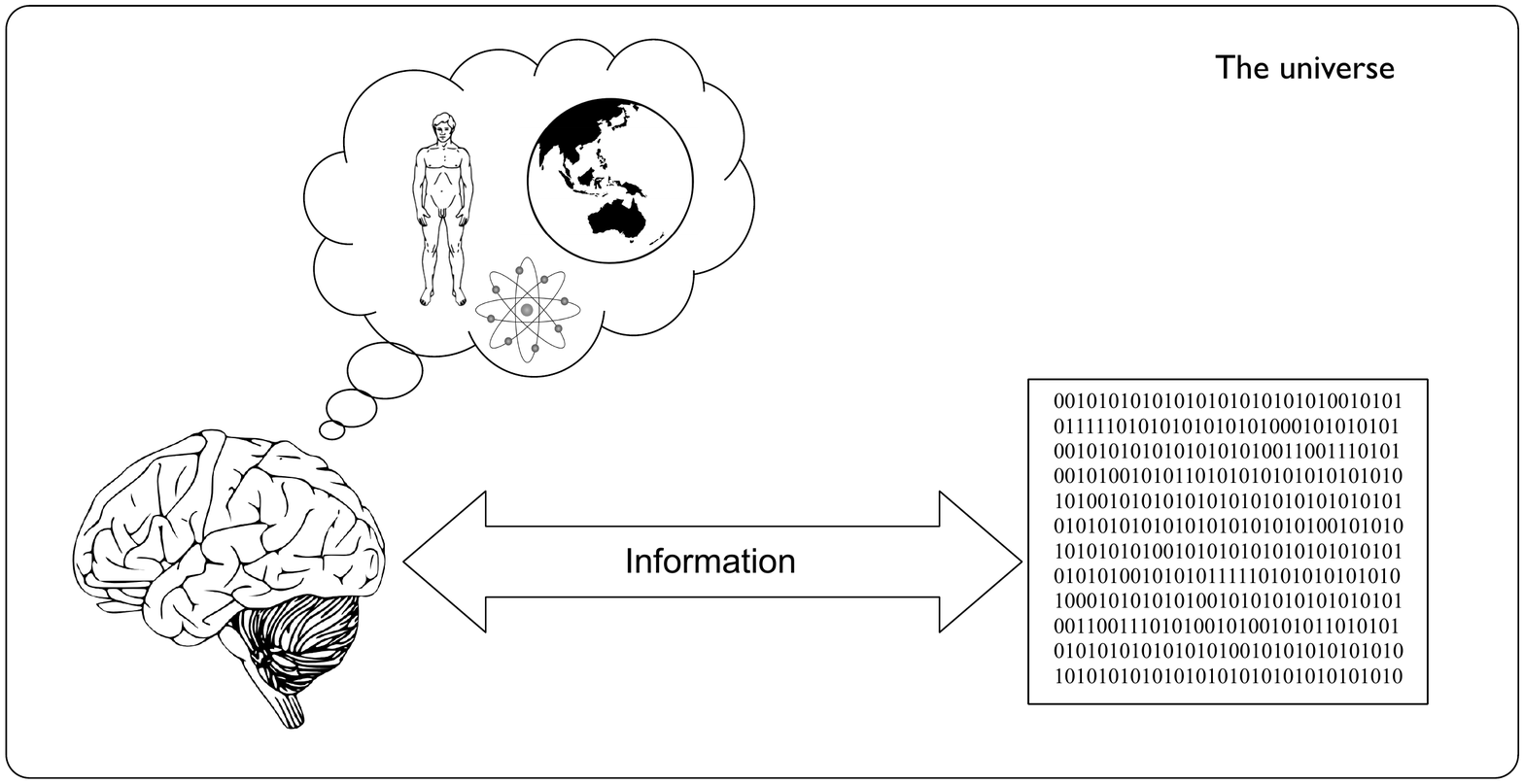}
\vspace{-1.5cm}

\caption{The brain creates the model of reality through  information exchange (in and out) with the `outside''. In this case we show the full universe sub-divided into a brain interacting with the rest of the universe by bi-directional information exchange, which physically occurs through  sensors and effectors.\label{machines} }
\end{figure*}
 
\section{Cognition from information}
 Consider then a universe and a brain in it, and the following two premises: 
  1) all human experience is generated in the brain, 2) the brain has  only access to information.  
If  we think of this brain as a modeling tool exchanging information with the rest of the  universe it is part of (the `environment'), then it follows that all questions about `reality' should be framed in an information theoretic framework with neuroscience at the center---an assertion that affects both neuroscience and physics.
By `environment'  here I  mean that which the brain exchanges information with, in the form  of sensory inputs   and the  outputs associated to what we call `actions'.  
 In truth, this brain can only state that  there is  information and information flow---see Figure~\ref{machines}.
 Information is the currency the brain uses to give states an identity and infer things about them.  In our  information-theoretic approach we could define a brain as a {\em a semi-isolated physical/computational system capable of controlling some of its couplings/information interfaces with the rest of the universe.   } This control of bi-directional information flow seems to be a key requirement for a cognitive system.

If all brains have access to is information, we can also think of brains as `information processing machines'---computers. Brains  work hard to process  the incoming information flow gathered by our senses and make decisions (actions). Brains send commands to effectors and harvest information from sensors.   Moreover, since the information gathered from our senses depends on how we choose to extract it from the outside world (through the passive and active aspects of sensing and modality choices), our model of reality must include a model of ourselves---of our `bodies' and internal `algorithms'. We must have a description of our bodies and of our acquired prior knowledge. These self-models are what we commonly call body representation and self-awareness.

I will posit here that as a modeling tool  the brain is organized in a hierarchical way. Sensors extract features and those are fused in different stages. The resulting higher-level features are again compressed, and so on. Moreover, high level outputs may affect low level compressors as well. Sensing is active, and  it is known that high-level processes can modulate perception. Hierarchical modeling and processing has been discussed before, see e.g., \cite{Friston:2003aa}, and here it will appear naturally in the application contexts described in section~\ref{sec:applied}.
 
  From an information-theoretic perspective, the result of a brain's interaction with its environment is to change both states, and this is mediated by an exchange of information. Information theory provides the conceptual framework for understanding the relation of brains with the universe.  Some of the brain-environment  information exchange is  mediated  by `controlled information membranes' (sensors and effectors), but some it is not and can be dangerous. In an eventual physical-informational theory yet to be developed---physicist John Wheeler's {\em It from bit} \cite{Misner:2009aa}---one would say that the information conveyed  by a bullet as it traverses a brain will change its state sufficiently to destroy its function as a modeling tool. 
   
   Given the above two premises, we can ask how brains infer what is `out there'. To survive--to maintain homeostasis---brains build models (algorithms)  to function effectively and to predict the future (i.e., future information streams). Thus, the first important observation is that `reality'  is a construct of our brains---a model. Reality is the algorithm brains build to sustain existence (and dodge information bullets). 
Examples of the models built by brains  and that are successful at compressing information include: space, time, charge, the dimension of space, mass, energy, atoms, quarks, tigers, people....  At this stage we can only point out that having access to a good model of reality, in terms of its compressive, operative and predictive power, may be  important for survival. I discuss in more depth the relevance of simplicity in the next sections.
 
 An important aspect  in coupling modeling with function in cognitive systems is the concept of  `pain' (or its antonym `pleasure'), which provides, perhaps again in a hierarchical manner, the objective function to optimize in  reality model building. Pain can be related to physiological concepts such as homeostasis. Modeling would appear to be necessary for homeostasis and therefore life.
 
To provide this information-centric model a modern incarnation  and illustrate the connection of modeling with function, consider the mapping of both environment and a into a computer program. In principle, a computer (a quantum one, to be precise) could simulate reality in as much detail as needed. In a practical implementation, we could consider the human as a program module with some interface with the other program (the surrounding universe). The program `human' could  for example instantiate a command to move its eyes. Some underlying routines would control eye gaze automatically as a function of head orientation through a vestibular submodule.  Such routines would implicitly contain a model of the humans' body and environment physics (e.g., the law of gravity).

 Let us clarify here that when we have only limited access to all the variables of a problem, the correct natural formalism is to work with so-called probabilities. Probabilities subsume and extend the concept of classical deterministic models. E.g., our classical notion of model is assigned to the expectation value associated to the probability function. This is actually mandatory in quantum mechanics, where models can only provide probability amplitudes.  I emphasize here the subjective stance in this paper. With Jaynes \cite{Jaynes:2003aa}, we refer to probabilities as a mental mathematical construct that represents in an unbiased way our knowledge of reality: a model.
 
\section{What is computation?}
Since  modeling and compression require a notion of computation, let us briefly analyze   this apparently well-defined notion. A simple mechanistic model of a universe computing its future is a closed system of billiard balls. Billiard ball computers have been proposed for the theoretical analysis of unconventional conservative computing \cite{Fredkin:1982aa}.  A subset of  balls can be arbitrarily identified as the Turing state register and the rest as the tape,  with newtonian mechanics as the transition function.   The model may seem overly simplistic, but it contains the most fundamental element in the theory of computation: dynamics. In an information theory framework, physical states are mapped into information and dynamics to computation. 

Similarly, the `real' universe evolves and in some sense computes its future \cite{Lloyd:2002aa}---and bodies and brains appear to be  part of it. However, this model already presents a puzzle. What defines the boundary between `me' and the `rest of the universe'?   From a fundamental physical viewpoint, it is not simple to define it.  That is, while boundaries may be defined in arbitrary ways, there does not appear to be a natural one.  As an interesting option, mutual algorithmic information  has been proposed as a means to define the natural physical boundaries of an organism \cite{Chaitin:1991aa}. A complete theory should be able account for this divide or do away with it.
Is there a guiding principle to separate `inference machines' from the rest of the universe?  Is the complement of an inference machine also an inference machine? 

\begin{figure*}[t!]

\centering
\includegraphics[width=10cm]{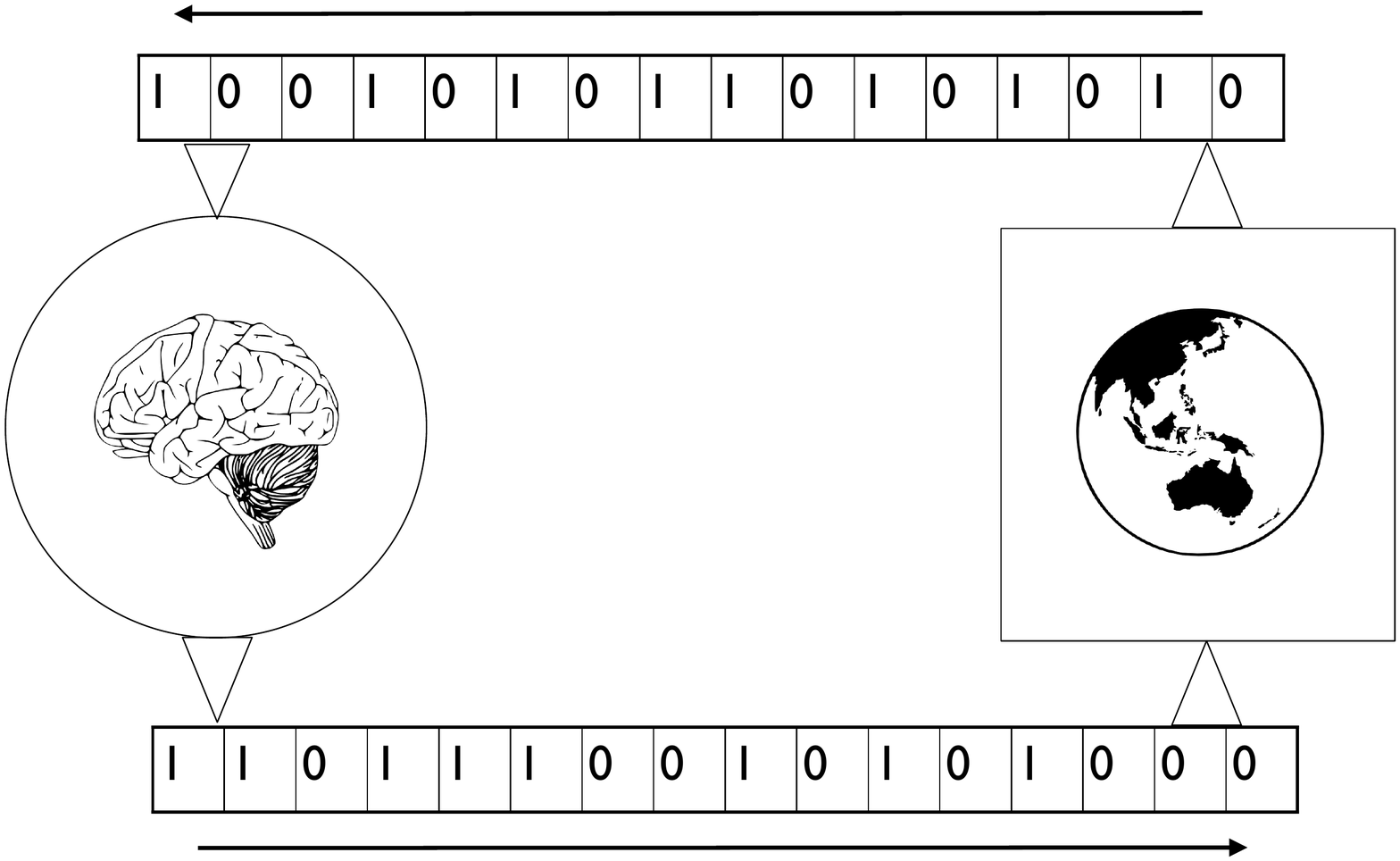}
\caption{The interaction of brain with the environment can be described by the symmetric coupling of two Turing machines.\label{coupledturing}}
\end{figure*}

A possible formalization of the problem may be the description of  the brain-universe relationship  by two coupled Turing machines. The concept of a coupled (`beyond-Turing') Turing machine has been described before  \cite{Copeland:1999aa}. A coupled Turing machine is one which can accept external inputs while operating, and which may also possess output channels. What we are proposing here is a specific symmetric construct involving two such machines: one in which the output of one  is the input of the other---see Figure~\ref{coupledturing}.  Alternatively, the theory of inference machines may be used \cite{Wolpert:2008aa} (further discussed below) and extended to consider the coupled case.

 It is important here to emphasize that computation is a dynamical phenomenon that requires recursion and  a concept of time.   Yet time may not be a fundamental physical concept. It may itself be thought of as a derived concept---another model (the `great simplifier' in the words of physicist Julian Barbour \cite{Barbour:1999aa}). Note that our billiard ball `universe-computer' can be described in 
a timeless way  by the  constants of motion (or, more directly, by the Hamiltonian of the system and the initial conditions). Can we construct a new information theory that does not rely on the concept of  recursivity and hence time?   It should be clear that in some sense computation does not add anything to the information of the initial state: everything is already in the program plus data which is available at time zero. Both time and computation are probably themselves models. 


%

\section{Simplicity}
  As has been set forth, what we call  reality is an interpreted abstraction---a mental construct---based on the observed regularities in multiple sensory input streams and in our own interactions (output streams). Hence,  the notion of reality is a model, with rules,  regularities  or invariants as its building blocks. We can easily show that physical concepts such as mass and space-time, or mathematical notions such as set and  even number are models we use to simplify  sensory information streams, typically in the form of invariants.  Compression is closely related to model or representation building, identification of invariants, memory and prediction.   The notion of repetition or recursivity is in this context the fundamental building block---algorithms require recursion. 
  
  In physical terms, simplicity, compression and modeling are all related to the identification of invariances in data streams. This is easily seen in the case of time-invariances:  once dynamical laws are found, the evolution of the system can be described through its constants of the motion, and these reflect time-invariants. In fact, solving the equations of motion is equivalent to finding all the constants of motion.  For example, in physical systems with time-translation invariance the quantity called energy is a constant of the motion, a conserved quantity. If invariants are found, e.g., empirically, the representation of the dynamics can be compressed.

  A general relation of conserved quantities to  symmetries in the physical world was formalized by Emmy N\"oether \cite{Noether:1918aa}, who showed that to each symmetry there corresponds a conservation law (see e.g., \cite{Goldstein:1980aa,Falter:1996aa}).  In our brain-centric context, symmetries refer to transformations of a multivariate input/output  stream at a given moment of time (e.g., how a rotation of propioceptive space affects vision and sound), or to more general transformation involving both space and time (Lorentz transformations). In essence, the existence of invariant quantities (conserved quantities) and symmetries that can be constructed from observed data points to non-randomness, and therefore to the possibility of compression and simplicity. In terms of a Lagrangian description of dynamics, if there is a continuous symmetry then there is a missing reference to a coordinate which is therefore conserved.
  
 In the following section I describe different approaches to quantify simplicity and their relation. Using standard modern language, complexity will be defined.

 
 
 \subsection{Simplicity from algorithmic information theory}
  Compression and therefore simplicity were first successfully formalized by the notion of algorithmic complexity or Kolmogorov complexity (KC\footnote{Kolmogorov complexity is also known as `algorithmic information', `algorithmic entropy', `Kolmogorov-Chaitin complexity', `descriptional complexity', `shortest program length' and `algorithmic randomness'.}), a  mathematical concept co-discovered during the second half of the 20th century by Solomonoff, Kolmogorov and Chaitin and  which provides a well-established albeit formal cornerstone to address the question of compression in brains---both natural or artificial. 
  We recall  its definition: the Kolmogorov complexity of a data set is the length of the shortest program capable of generating it.  
More precisely, let $U$ be a universal computer (a Turing machine), and let $p$ be a program (see e.g., \cite{Cover:1991aa,Li:2008aa}). Then the Kolmogorov or algorithmic complexity of a string $x$  with respect to $U$ is defined by
\begin{equation}
K_U(x)= \min_{ p: \, U(p)=x} \, l(p),
\end{equation} 
  the minimum length over all programs that print  the string $x$ and then halt. Although a technical point, the restriction to programs that halt is important, for no program is then the concatenation of other programs. 
   An important fact is that this is a  meaningful definition: although the precise length of the minimizing program depends on the programming language used, it does so only up to a constant.  That is, if $U$ is a  universal computer, then for any computer $A$ we can easily show that $K_U(x) <  K_A(x)+ c$. The constant $c$ is the length of the program for $U$ to emulate $A$.

Unfortunately, G\"odel's incompleteness theorem, or its equivalent, Turing's halting theorem,  implies we cannot compute in general the KC of an arbitrary string:  it is impossible to test all possible algorithms smaller than the size of the string to compress, since we have no assurance of that will ever halt \cite{Chaitin:1995aa}.  However, if we change the rules of the game and state that there is a finite  computation time, compressibility  becomes a practical question. A theory of `practical' complexity    can tell us what the expected length of the shortest programs can be if computational resources are finite (in time and space), or if there are constraints in natural compressing mechanisms.

A further connection to simplicity using the concept of KC was developed by Solomonoff \cite{Li:2008aa} with an emphasis on statistics and prediction.  The fundamental quantity here is the algorithmic or universal (un-normalized) probability $P_U(x)$ of a string $x$. This is the probability that a given string $x$ could be generated by a random program. An important result is that this is given by 
\begin{equation}
P_U(x)= \sum_{p: \,  U(p)=x} 2^{-l(p)}.
\end{equation}
Here the earlier remark that concatenations of programs cannot be new programs is crucial.  This result says that short programs contribute more to the probability of observing a given data string. A further important result connecting this probability to KC is that 
\begin{equation}
P_U(x) \approx 2^{-K_U(x)},
\end{equation}
that is, the probability of a given string to be produced by a random program is dominated by its Kolmogorov complexity.  What this result says is that the probability of observing a string is actually dominated by the shortest program capable of generating it. More precisely, suppose that we are given a string $x$ (a measurement). Then we can ask what is the relative probability of observing a future string $y$, i.e., the total string $xy$ (concatenation implied), or the alternative $xz$. This will be given in this framework by
\begin{equation}  
{ P_U(xy) \over  P_U(xz) } = {\displaystyle \sum_{p: \,  U(p)=xy} 2^{-l(p)} \over  \displaystyle  \sum_{p: \,  U(p)=xz} 2^{-l(p)}  } \approx 2^{ -[K_U(xy)- K_U(xz)]}. 
\end{equation}
An important aspect  in this approach is that predictions are made considering all the appropriate explanations with their proper weights, and not only the most likely one. 

The precept that short explanations are in some sense  more likely is also the essence of the Minimum Description Length (MDL) and the  Minimum Message Length (MML)  approaches to statistical inference \cite{Li:2008aa,Wallace:1999aa,Vitanyi:2000aa}.  Among possible explanations (programs) for data (an observed data string), those which are shortest are more likely.

Two points need to be clarified to highlight the connection of MDL with KC. The first is that in general we can think of the shortest program as the sum of two sub-programs: one that encodes the regularities in the data, and the other which gives the error or remainder.   This is rather intuitive.
If models do not fit the data well, they will have to keep an explicit, uncompressed tally of the error as part of the program. Thus, the concept of misfitting  appears here in a natural way, providing the connection to Bayesian estimation---further discussed below. The error term in this framework provides a solid definition for the concept of `noise', which is simply that part of the data we have no power to model with the information we have access to (if we have prior information it should be packed with the new one).  For example, we could think of the concept of `sphere' as arising from MDL analysis of all haptic experiences, as a sort of average shape---the simplest archetype for experienced objects.

Thus, in ideal MDL shortness is measured in algorithmic complexity terms, and this approach is essentially equivalent to KC minimization. Model plus error separation happens in a very natural manner in the context of compression. That is, if a set of data is to be compressed and we have no other information, KC or ideal MDL provide the recipe of shortest program length, which gives, as a byproduct, a recipe for separating something to be called ``abstract model" from ``noise". Similarly,  in Bayesian terms code length of the model and code length of model plus data together correspond to prior probability and marginal likelihood respectively in the Bayesian framework \cite{MacKay:2003aa}. 

The second is that in practice we know that KC is in general uncomputable. For this reason, in MML  a particular compression scheme based on Shannon information theory using efficient codes is used. This method states that given some data and an efficient ``code" to represent models and data, the best model is that which minimizes the sum of the length (in bits) of the model plus the length of the data once encoded by the model (which we can think of as the error or noise we cannot model in Bayesian terms). This can be shown to be equivalent to selecting the  highest Bayesian posterior probability, which we discuss in the next section. 
The use of efficient coding, based on entropy, as opposed to algorithmic coding, based on complexity,  implies that this approach is practical, since in practice MML can be computed. Moreover, it can also be shown that in a statistical average sense, entropy is a good proxy for complexity \cite{Cover:1991aa}. 

It can also be stated that the only difference between practical MML and its universal version in ideal MDL is that the second one uses a universal Turing machine, whereas the former does not in the interest of practical use.




\subsection{Statistical inference: Bayes and Occam's razor}

Jaynes's formalization of probability theory \cite{Jaynes:2003aa} is   closely related to the philosophy of this paper, since it  too takes a subjective stance. Jaynes discusses how a cognitive system (a human brain or a robot) would seek to formalize logic and its extension, probability theory, in order to optimize the search for models.  In particular, Jaynes defines probability theory as a natural extension of logic, targeting the problem of optimal inference from models: reasoning about propositions and their plausibility, which are tools for model building. 
We recall here the  conceptual kernel of Bayes theory is the relation for the `probability'
\begin{equation}
p(a,b)=p(a|b)\, p(b) = p(b|a)\, p(a) 
\end{equation}
where $a$ and $b$ may describe for example data and models. 

The relation of Occam's razor to Bayesian theory is discussed for example in  \cite{Jaynes:2003aa, MacKay:2003aa, Tipping:2004aa}.  Given two models, $b_1$ and $b_2$, their relative probability given some data $a$ is
\begin{equation}
{ p(b_1|a) \over p(b_2|a) } = {     p(b_1) \, p(a|b_1)  \over  p(b_2) \, p(a|b_2)  } .
\end{equation}
Without access to any prior information, we may say that the two models have equal priors, and conclude 
\begin{equation}
{ p(b_1|a) \over p(b_2|a) } \sim {       p(a|b_1)  \over    p(a|b_2)  }.
\end{equation}
In Bayesian theory the quantity $ p(a|b) $, with $a$ referring to data and $b$ to a hypothesis or model is called the {\em evidence}. It is a measure of how well the data is fit by the model. Without prior information about the model probability distribution, the probability for a model is thus proportional to the evidence. The evidence gives the probability to observe data if the model is true. It is proportional to the goodness of fit but inversely proportional to the model available search space---the model volume. For example, a theory with two parameters will provide a higher evidence than a three parameter model if goodness of fit is similar.   In essence,  but rather loosely, goodness of fit being equal and with a uniform prior, simpler models will yield greater probability for the data, because they live in a smaller space.  


The use of uniform priors was already proposed by Laplace. Laplace proposed a related concept called the `principle of indifference'. This principle states that given a set of possible scenarios about which we have no information, the only logical thing to do which will represent correctly our knowledge is to assign to each equal probabilities.  More precisely, suppose that there are $n > 1$ mutually exclusive and collectively exhaustive possibilities. The principle states that if the $n$ possibilities are indistinguishable except for their names, then each possibility should be assigned a probability equal to $1/n$. Assigning equal probabilities to a set of scenarios amounts to providing the simplest model of the phenomena. In this case the model is a probability function, and a standard measure of probability of is entropy. This principle was formalized  by Jaynes as the so-called Maximum Entropy Principle  \cite{Jaynes:2003aa,Jaynes:1957aa}.

We end this section with a word of caution. Despite statements to the contrary, the above arguments do not prove the validity of Occam's razor or simplicity as a principle for inductive inference.  
As pointed out in \cite{Wolpert:1995aa} and \cite{Domingos:1999aa}, the no free lunch theorems for optimization \cite{Wolpert:1997aa} imply  that unless some assumptions are made about the space of possible algorithms and the space of data, Occam's razor does not follow.  For example, the connection of Bayesianism and Minimal Description Length can be justified if the universal prior of Solomonoff applies, but not  in general  \cite{Vitanyi:2000aa}.  This is a very important point, and implies that we need to search for explanations for the importance of simplicity in the realm of biology and neuroscience, and not in mathematics alone. We provide some ideas in section~\ref{sec:importance}. 

\subsection{Inference machines}
For completeness we briefly mention a new approach to complexity. An absolute measure of complexity using the concept of (strong) inference machines has recently been proposed \cite{Wolpert:2008aa}. Inference machines theory (IMT) is a mathematical formalization of physical devices performing observation, prediction or recollection tasks---what we call here cognitive systems---and they generalize Turing machines. An important aspect in the theory is that an  inference machine is itself part of the universe it is trying to infer. This recent formalization has already delivered very interesting results, including the fact that for any inference device there exist functions it cannot infer, and that there can  only be one strong inference device.

A natural definition of complexity is provided in this formalism in terms of the size (in some sense) of the input subspace of universes allowed by the `program'  for the inference device to produce the required output data stream: the bigger this subspace, the simpler the program is.  In effect, the setup of the inference device constrains the types of universes in which the computation takes place. A simple program is less constraining.  What is especially interesting about this theory is that there is a notion of an absolute complexity (defined by the unique strong inference device), as opposed to the standard computational theory, where complexity is to some extent relative.  Further work is needed to understand the implications of IMT  to the subject of this paper.

\section{Why is simplicity important?\label{sec:importance}}
As we have seen, simplicity is clearly useful in compression, but its applicability to inference is not obvious from first principles. 
If the universe and our world are really simple, as posited by Leibniz and others, then simple models  should be sought.  As discussed above, simple models can be said to be statistically more likely in some sense: if `god' is a monkey typing random programs on a computer (i.e., if Solmonoff's universal prior applies), then simple programs are more likely explanations or causes of observed data and simplicity will not only be practical, but also better at prediction. Unfortunately, it has not been possible to justify this assumption  yet. However, simplicity offers advantages even in the absence of  a simple universe.  We can consider different aspects of this question.

Firstly, I note that brains, as inference machines, need to survive in their environment. We should therefore analyze evolution and natural selection as the conceptual basis for the discussion. 

Cognitive systems carry out three main different tasks, which we can classify as recalling, observing and acting, and predicting or extrapolating---corresponding to past, present and future. In all three of these tasks---also discussed by \cite{Wolpert:2008aa} in the context of IMT---we can find value in simplicity. 

{\bf Past:} memory is definitely an important dimension in the discussion.  Simplicity  is valuable as it allows to store information more economically, using less memory resources---although at the expense of computation. If a simple model is found to compress the data, it will become easier to store and eventually recall. Good  models can account for  large data streams, and therefore provide a deeper view into the past. Moreover, as explained by Jaynes, simple models represent knowledge in a fair way. Suppose axioms A, B and C `explain' (decompress to)   the facts.
We could also add another axiom, call it D, if it does not conflict with the facts. But adding it is a disservice to our representation of knowledge, because it is adding unfounded aspects to the model, adding information that is unwarranted and potentially biased---an invention.

{\bf Present:} simplicity is advantageous in deducing, planning, acting or deciding, including in observation, experimentation and decision---i.e., dealing with the present.  Simple models lead to simple deductions, with less clutter from bogus tenets leading to `model noise' (such as those that may derive from axiom D above). Simple models are easier to run, requiring less memory resources and memory access. Experiments provide the information we want with less clutter from spurious variables to be controlled if they are easy to execute. I emphasize that the act of experimenting is crucial for brains, as it is through experimentation that the world is explored. When our hands reach out to touch an object, we are carrying a multi-modal experiment to build, based on earlier models, an improved model or representation of the object.  The existence of an underlying simple models allows for the design of simple experiments.  Likewise, a simple model should be easier to use for decision making. Given the No Free Lunch theorem, using simpler models appears to offer only advantages, even if the universe is not simple. That is, if there are no generally superior algorithms, using the simplest available is a practical strategy.

We may also add here two obvious but important observations. First, in the act of observation brains access only partial information from the environment. This gives rise to the notion of noise, for example. And it also entails the notion of `coarse graining', which may be a key feature of living beings.  Coarse graining means that special combinations of data are observed (e.g., macroscopic averages). For example, we can easily measure the center of mass of a planet (or at least, much easier than the position of each atom). It also happens that we can model and predict special coarse grained quantities, while it would be impossible to predict every detail of the environment. As  an example of this we refer the reader to  \cite{israeli:2004aa}, where it is shown that it is possible to coarse grain irreducible (incompressible) cellular automata in order to make them compressible. The  Kolmogorov complexity of an incompressible stream is basically the length of the data stream.
Incompressibility here means that in order to model what will happen in the future, there is no other approach than brute force computation of every single step, there are no regularities to exploit or shortcuts to take.  
 Secondly, the act of observation perturbs the universe, because information is injected into it. Observation or sensing are active and disturb that which we observe. Both of these statements are true in classical physics but take on an extreme form in quantum physics. These obvious points lie at the core of the foundations of physics, a point we come back to below.

{\bf Future:} from the prism of evolution and biology, prediction or extrapolation are certainly an important element for survival, and  simplicity is  advantageous for prediction tasks as well.  As we have discussed, simple models are less biased \cite{Jaynes:2003aa,Jaynes:1957aa}, representing in truthful way what we know from the available data and using the least resources.  In this paper I take the view that, given the No Free Lunch theorem, simple models are the most practical---in principle as good as any other for prediction purposes. Simple programs are easier to use for prediction, requiring less in terms of memory resources to store. Perhaps this is the true evolutionary origin of simplicity, as the original simple organisms counted on few resources and could only construct simple models. Simplicity through recursivity allows the replacement of  memory space for time. This logic may also apply to DNA. Evolution would seem to lead to a layers of simplifying processors.  Finally, short programs are also easier to debug and correct, which is certainly important for biological processors such as DNA.

The relevance of simplicity for extrapolation is also at the core of the question of why mathematics and simple theories turn out to be the right ones in science, and especially in physics---a mystery that has perplexed many thinkers, including Einstein and Feynman. A possible answer is that inference machines will by nature---including limited resources---always seek and find some simplicity, even in random data. Let us recall here that any desired sequence can be found in a truly random number. An inference machine exposed to a lucky (compressible) portion of the stream will deduce all sorts of things and be fooled by the passing mirage. This is the complexity, or rather simplicity, version of the Anthropic principle (see e.g., \cite{Susskind:2005aa} for a discussion of Anthropic theories). The ``inferotropic principle" would state that  without simplicity there would not be infererence machines.  An related explanation may be provided by coarse-graining. Perhaps the first task of inference machines is coarse-graining for compressibility: choosing correctly how to observe and interact with the world  may be the key to simplicity. This would give an added dimension to the statement that brains construct reality.

\section{Simplicity and cognitive science\label{sec:applied}}

\subsection{Fundamental neuroscience}
Different experimental scenarios could explore the hypothesis that the brain is a compression machine. One especially interesting approach is to study  how compression and model building are  handled  at different hierarchy levels.

As is well known, there are mechanisms in the brain that are well adapted to modeling patterns and detecting change. Some of the related  experimental work  involves studying the response of the brain to different input sequences. It is possible today to provide  stimuli with accurate timing to the brain (visual, auditory, haptic) and then measure the response as seen using scalp electrophysiology. For a brief review of compression in visual processing, see \cite{Chater:2003aa}. In audition, it is typically necessary to average the response over several hundreds or thousands of such events in order to filter out the noise  and observe what is called an Event Related Response (ERP). 

In Mismatch Negativity experiments (MMN),   auditive {\em patterns} are presented to the subject and the response of the brain is measured using an EEG or MEG recording system  \cite{Naatanen:1978aa,Grau:1998aa}. The MMN component is thought to result from the comparison process between the  stimulus and  the neural representation  of the auditory past maintained in  sensory memory \cite{Atienza:2003aa}.  This could be a simple recording of the recent past, or an abstracted model (`abstract invariance' in the language of \cite{Picton:2000aa}). MMN is a low level automatic response generated by infrequence pattern breaks,   not requiring attention and present even during sleep. It has been used extensively to study the perceptual organization of sound as well as memory. It has been shown that MMN mechanisms can detect regularities and use them for prediction. The appearance of a MMN wave indicates that the system has made a wrong prediction, or, equivalently, that the current model needs updating. Therefore,  MMN can be used to study how the central auditory system encodes auditive time series near the low end of the processing chain. 

MMN experiments in effect measure the change in response when the inferred patterns are broken. The phenomenon of MMN basically illustrates the fact that the brain, at some very primitive level, is able to extract patterns---to compress information. Although algorithmic complexity theory has not been applied to this problem, it seems a very interesting prospect. We can for example measure how many iterations are needed before the pattern is assimilated as a function of algorithmic complexity, and we can try to understand the type of pattern that the brain's mechanisms involved are best at locking into.  

 In more detail, when a {\em deviant} tone is presented, the ERP differs from  the {\em standard} response. This illustrates the fact that the brain is acting like a real time modeling engine. What makes this especially interesting is the fact that the level of response to the deviant may be related to the complexity of the sequence. 
In the experiments described in \cite{Atienza:2003aa}, the following  repeating (non-random) sequence was tested: 

\bigskip

{\small

S1: ABABAB  - ABABAB  - {\bf B}ABABA -  BABABA -  {\bf A}BABAB -  ABABAB ... }\smallskip

\noindent 	  where A and B stand for two tones and  the dashes for silence periods. Each group of letters is called a `train'. If the silence periods were sufficiently short, a MMN event occurred at each tone repetition (marked in bold). On the other hand, for long silence periods (inter-train interval $>$ 240 ms), there was no such MMN response. This shows that the relevant brain's auditory subsystem  treats each train independently when they are sufficiently separated in time, and  as a continuous stream otherwise. Let us analyze this in more detail.  We can code this sequence as
$$ 
S1= R_0(R_3(AB)+R_n(-)+R_3(BA)+R_n(-))
$$
using a simple coding language where $R_n$ stands for `repeat $n$ times', with $n=0$ referring to  infinite repetition.  In the case of short silence periods, the stream can be rather efficiently approximated using a very simple first order rule, namely  
 $$
 S_{n+1}=NOT(S_n)
 $$
which we can loosely call `alternate' symbols. Then, the continuos sequence \bigskip

\hspace{0cm} {
S1: ABABABABABAB{\bf B}ABABABABABA{\bf A}BABABABABAB ... } \medskip

\noindent results in a MMN coding of \bigskip

\hspace{0cm} {
MMN[S1]: 000000000000{\bf 1}000000000000{\bf 1}000000000000 ... } \bigskip

\noindent This coding scheme results in some errors (represented here by 1s) occurring at regular intervals, which are however in principle easy to code by a subsequent compressing layer. So MMN acts as a  low-level compressing or coding system. A hierarchy of such coding systems could detect and encode regularities at different time scales. If there is a hierarchical complexity scheme in the data, an appropriate pattern detector can be constructed from simpler ones by concatenation of simple pattern detectors---but other possibilities probably exist. The existence of such structure in the natural data would explain the fact that simplicity is preferred, and why low level simplicity couples to more automatic detection systems. This process may also be the essence of music. 

What happens if the inter-train intervals (ITIs) become sufficiently long? Clearly, the simple alternating rule cannot account for long-scale regularities (recall that the sequence in this particular experiment was fully predictable).  In \cite{Atienza:2003aa} it was shown that no MMN response is elicited with long ITIs, indicating that either a)  the coding scheme captured all the complexity  (perfect prediction), or b) the MMN mechanism was partially or totally inactive (the rule too complex to learn in its entirety).  In order to test the first hypothesis some tests using pattern breaking at the larger scale (e.g., repeating 3 times one of the trains randomly) could be carried out (but unfortunately they were not in \cite{Atienza:2003aa}). However, the results in \cite{Atienza:2003aa} using  another sequence (see below), suggest that such larger scale rules are not encoded. Experiments in with ITIs and non-random series included inserting repetitions at other places, e.g., \bigskip

\hspace{1cm} {\centering \small
S2: ABABAB  - -   A{\bf A}BABA   - - BABABA    - - B{\bf B}ABAB  ... } \bigskip

\noindent again in predictable manner. What was observed is that the inferred pattern by the auditory system was again  that of alternation. With long ITIs, this rule generated no response in case  S1, but did in case S2 and similar others.    Moreover, the results also hinted at the  existence of higher level encodings or `logical consequences' of the simple alternating rule. For example, MMN was also observed to occur when the consequence $ S_{n+2}=S_n$ is violated (experimentally, in S2 after onset of the third tone).  
A rather logical interpretation here is an intended abstraction in the coding scheme, some sort of relativity principle. The abstract rule of alternation could be seen to be `coordinate independent', e.g., indifferent to how it is started, ABABAB $\sim$ BABABA. This would be a stronger form of simplicity encoding. In fact, this would be the best compressing method if the system is handling trains as independent events, or, e.g., if the alternation of trains ABABAB and BABABA were truly random.


Although the study in \cite{Atienza:2003aa} was not explicitly designed for complexity analysis, 
some could easily be devised using random violation of rules. Consider the patterns \bigskip

{\small 
\hspace{2cm}  Z1= ABABAB - -  ABABAB  - - ABABAB ...

\hspace{2cm}  Z2= AABABA  - -  AABABA  - -  AABABA ...

\hspace{2cm}  Z3= BAABAB  - -  BAABAB - -  BAABAB ...
}\bigskip

These sequences have been ordered according to complexity. E.g., using Unix's {\em compress}  (which uses LZW data compression) compresses most the first, then the second, then the third sequences.  The first sequence is efficiently described by the AB subunit and the silent space (common to all). To describe the second one, we need A twice  and then BA twice plus the space. The last sequence requires also two subsequences, BA and AB, the first one repeated once, the second twice, 
and Z1$=  R_0(R_3(AB)+R_n(-))$, Z2 $= R_0(R_2(A)+R_2(BA)+R_n(-))$ and Z3$=R_0(R_1(BA)+R_2(AB)+R_n(-)) $.
We can see how the length of the programs grows. 

Targeted  MMN experiments could thus shed light on the role and strategies for compression and simplicity in memory and prediction. We would expect that complex patterns might elicit a weaker or delayed MMN response (in agreement with \cite{Atienza:2003aa} and other experiments, e.g., \cite{Kanoh:2004aa}) and    take longer to learn (for example, \cite{Benidixen:2008aa} discusses how rapidly simple but abstract rules are learned). Or we might find that  MMN related mechanisms can code only very simple rules, leaving algorithmic integration to higher level processes which are more difficult to observe using EEG.  Finally, the MMN paradigm, which is used aurally and apparently engaging low level pattern detection systems, could be expanded to other modalities (somatosensory, visual, etc.) using emerging virtual reality (VR) technologies, as well as in the complexity scale to explore model building in different human cognitive sub-systems.


\subsection{Presence}
The idea of model simplicity can be used in the context of the  field called {Presence}. 
In line with the theme of this paper, Presence can be said to study how the human brain constructs the model of reality (including body and self)  through  full (immersive) or partial (mixed) digital interaction. As such it is part of a wider family of disciplines studying how cognitive systems build models of their environment and interact with it, i.e., cognitive sciences.
 In Presence, the technological goal is to place a brain in a controlled, convincing and interactive information bath, so that subjects feel and act as if the designed experience were real \cite{Slater:2009aa}. Presence provides the natural scientific approach to study the issues discussed in this paper at different perceptual and cognitive levels.
The designed information bath can be fully digital (immersive), or mixed, containing both real and digital elements. 

Let us consider now the following thought experiment. A person is placed in an enviroment in which computers prepare and control   all sensorial inputs (via sophisticated  audio and video 3D displays, haptics, olfaction, vestibular stimulation, etc.) and in which the person's brain commands are intercepted at the level of, say,  peripheral nerves or even using brain-computer interfaces.  If this experiment is carried out successfully, the person will feel fully  immersed in the  virtual environment and, in measurable terms \cite{Slater:2009aa,Sanchez-Vives:2005aa, Harvey:2005aa, Slater:2007aa}, act as what he or she is experiencing were real. In this experiment, it suffices to describe the universe in terms of bits, because the environment is really created by  program in a computer managing  sensorial inputs and brain outputs---see Figure~\ref{machines}. 
We can identify three main elements in this setup: 1) a human or animal agent (a brain), 2)  bi-directional human-machine interfaces, 3)  a machine agent. All these are necessary to create the subject-environment loop.
This ideal experiment represents the central paradigm in the field. The reader familiar with the movie Matrix will easily recognize the concept of immersion, capture and replacement of efferent and afferent data streams, which in the movie are implemented in an invasive manner. 
 

Reality, according to this paper and illustrated by the above thought experiment, is a model arising from the coupling of brains and bits. As has been argued, among all the possible models that can account for our observations, all other things being equal the simplest ones are ranked more probable by the brain. 
 Following the logic laid out in this essay, I contend that in order to achieve `more'  Presence, simplicity in the explanation of what is happening both at low and high cognitive levels   is a key aspect. 
Here simplicity will include a consideration of  departure from models built from prior experience, which could be inherently complexity-increasing.  
 We state this as the simplicity hypothesis and as a virtual reality design principle: {\em given alternate models (interpretations) for a given mediated experience, a brain will select the simplest one it  can construct that agrees with the data and with prior models. Furthermore, the simpler the model, the more real the experience will feel.}  

Three important aspects need to be emphasized here. The first is that models must take into account all the available information, including past data encoded as a priori model distributions.  This is agreement with the algorithmic complexity or MDL description of modeling. Agreeing with prior models just means that in the search for simple models we will be treating all available information in a democratic way. 
The second is that in order to tie  these concepts with  modern Presence ideas, we  need to postulate the existence of a hierarchy of models, or what is the same, a hierarchical compression scheme going from low level perceptual mechanisms (sensorial spatio-temporal binding \cite{Harvey:2005aa}) to higher cognitive modeling (an extension of binding concepts to higher processing stages). In addition, different modeling hierarchies will have access to different amounts of working memory.  Third, the principle of simplicity needs to be applied with care, since we do not mean to say that the brain attains by any means the KC limit, but only a sufficient approximation of practical value.

Finally, if the brain's goal is to construct models for survival, and if pain---as related to homeostasis---provides the objective function to optimize through model building, then a rather direct road to Presence is through the stimulation of pain receptors---nociception. I come back to this point below, but note here that human skin is rich in pain receptors associated to mechanical pressure or deformation, chemistry and temperature.


\subsubsection*{Simplicity and sensorimotor consistency with models}

The relevance of modeling and simplicity in Presence  is already implicit in work  describing the importance of so-called `sensorimotor contingencies' and the binding problem. This refers to the process by which the brain effortlessly uses spatially and temporally segregated activity in neuronal ensembles to form a unified perception  \cite{Sanchez-Vives:2005aa,Harvey:2005aa} and to the importance of `coherence' of motor outputs and multimodal sensorial streams.  It should be clear at this point that sensorimotor contingencies  refer to the {\em consistency} of inputs and outputs with  models, and hence, to the idea of simplicity.  They could be better called, in the spirit of this paper, sensorimotor consistencies.

As an example, let us consider  the so-called `rubber arm' experiment \cite{Botvinick:1998aa,ijsselsteijn:2005aa}.   
In the experiment, the real forearm is hidden, and a fake one is displayed and stroked. While the fake arm is stroked, the real one is simultaneously stimulated but hidden from view. Given the sensorial visual and haptic evidence,  the experimental subject could hypothesize: 
\begin{itemize}
\item {\bf Model 1:} {\em That is my hand being stroked} -- This is a simple plausible theory (which we may call `body illusion') that may account well for what is presently happening to the subject, but which disregards the recent past as coded in high-level memory (``I am in an experiment"). Higher-level cognition  will not  accept this model, but low-level models in the brain  have no access to long-term memory and complex processing, so they are easily fooled by the experimental setup. The experimental evidence available to the subject can be accounted  for by this simple model and the subject will feel at some level that the fake hand is real, something which can be inferred through physiological measurements  of  galvanic skin response, for example. Of course, the better rendition of a fake arm we can provide, the less noise the subject will need to deal with to accept the illusion.  Moreover, if the subject were to suffer from some severe form of amnesia affecting short-term memory, this explanation might become more plausible at all levels and even become `reality' after a sufficiently long time.  
\item {\bf Model 2:}  {\em There is a complex set-up in place to fool me into believing that this is my hand} -- This is a more complex model, but in fact it is the simplest explanation available to high-level processing that is consistent with all the modeling hierarchies and takes into account all the available data to the subject (e.g., from birth). However, this model, despite being the simplest and most truthful  taking into consideration all the data, is not available to all processing subsystems---where it is not the simplest. Low-level modeling will still be swayed by the illusion.
\end{itemize}
 As can be seen,  a cross-hierarchy modeling conflict is taking place. If a hammer is suddenly driven into the fake arm, the subject will often pull away. This is not surprising, since such reflexes are controlled at low level. Thus, we cannot say that the first or the second models of reality are in control. Both are active in some sense.

Another similar example is provided by the so-called  `Pinocchio' illusion (see, e.g., \cite{Lackner:1988aa}). In a variant, a blindfolded subject is made to stroke a third party nose while his/hers is  simultaneously stroked by the experimenter. The coherence of inputs (haptic inputs through nose, hand and propioception) supports the ``I have a long nose'' proprioceptive low level body model. Again, a higher level explanation involving the cortex and an experimental conspiracy (``there is a set-up to fool me'') is more complex and simply unavailable to low level modeling systems. 


To summarize, we conjecture here that  the feeling of Presence, as measured by subjective or objective ways, is increased if the induced sensorial input (the input data stream), has a low complexity, i.e., it can be modelled in a simple manner by the subject's brain compressing subsystems. Physical consistency in the inputs, in the sense of there being a simplifying low level model available to match the data, is an important element to enhance Presence, as it connects with low level, small memory capacity  modeling mechanisms. As we progress higher in the modeling hierarchy, Bayesian prior expectations (memory  play an important aspect: explanations with a better match with past experiences are in a sense inherently simpler. 

 There can be conflicts across modeling hierarchies, and in some sense there is no unique model of reality, but several competing ones.  We cannot for the moment state which levels in the modeling hierarchy are favored when in disagreement, but it would seem to follow that  low-level modeling mechanisms will associate to  more primitive responses---and viceversa. We can also   conjecture that low-level (perceptual) modes and high-level (cognitive) models will reinforce the illusion if they are both in agreement.   Finally, the stimulation of nociceptors should stimulate (motivate) model building significantly, and may thus enhance Presence.
 
 For simplicity, in what follows we will refer to  only two modeling hierarchies, `low' and `high', but in reality we can expect that  there are many such processing layers.

\subsubsection*{PI and Psi}

The conceptual framework presented here is compatible with the recent proposal for concepts such as `Place Illusion' (PI or $\Pi$) and `Plausiblity' (Psi or $\Psi$)   to establish a theory of Presence \cite{Slater:2009aa}. Both of these find their roots in the theme of  model building, and in particular  on the consistency of interaction data streams with low level (PI) and high level (Psi) simple models.


 PI refers to a {\em qualia}:  {\em the strong illusion of being in a place in spite of the sure knowledge that you are not there}.  It is the space-analog of the body illusions (rubber-arm, long-nose) that we have discussed above. Virtual reality (VR) systems---e.g., using immersive head-mounted displays---are typically quite successful at generating this illusion. From the point of view of VR design, we could also add the requirement that the perceived place correspond to the intended one by the VR designer.  To achieve PI, the input/output streams must respect the structure of low level models that code things like ``if I initiate motor commands to rotate my head in such a way, the visual field shall rotate in such a manner". Such low level sensorimotor models are probably rather rigid. For example, based on a model of reality (body, space and physics), the vestibular system exerts direct influence on eye muscles in order to sustain stability of images in the retina (and hence simplify the resulting data stream).  Such models  encode body models and ``routine existential physics", including Newtonian physics, geometric optics, etc.,  but, e.g.,  not general relativity or quantum mechanics, which are beyond normal experience.  
 
 PI can be generated by providing this natural consistency between stimuli and actions, i.e., input and output streams. In order to produce it, immersion in an appropriate virtual reality environment is necessary. As explained in \cite{Slater:2009aa}, PI  relates to how the world is perceived (perception). The associated modeling layers are probably hard-coded genetically to a great extent (as opposed to learned). Similarly, we could extend this concept deal with how the body is perceived (BI or `body illusion) or more generally, to the physical world (PHI  or `physical illusion)

On the other hand, Plausibility (Psi) refers to consistency of the data stream with prior learned models. That is, it is associated to higher-level modeling tasks of the brain.  As defined in \cite{Slater:2009aa},  {\em Psi is the illusion that what is apparently happening is really happening 
(even though you know for sure that it is not).} Thus, it represents a step upward in the modeling hierarchy (but not to the top). From the point of experience design, we could also add that what is intended to appear to be happening is the actual perceived illusion. That is, what is experienced is actually the model of reality $M$ which the VR designer is trying to convey.  We discuss this point further below.

The ultimate level in this context may be called `Reality Illusion' (RI) and would occur if the  input stream is convincing at {\em all} modeling levels, even the highest ones. Today, the only way to achieve this would be to place somebody in a real but very controlled environment. This is the realm of con artists, scams, and related human affairs. It may also happen during dreams, perhaps because some of the model checking mechanisms are turned off. I discuss below an `almost real' scenario.

It should be apparent that  VR technology will become  a very powerful tool to study human modeling hierarchies.
I  provide now a more mathematical description of these ideas using Bayesian theory and algorithmic complexity. For a related application of Bayesian theory to neural coding and computation see, e.g.,  \cite{Knill:2004aa}. 

 First let us state what a {`reality model'} or hypothesis  is here. A model $M$ is a program running in a cognitive  system (or in a cognitive system)---a finite binary string---that can   compress and predict the information associated to a set of events. To be precise, let us  define the set of  events  $E$ as the input/output information stored in the short term working memory of the subject---another finite binary string. Event streams contain information about input sensorial streams, output effector streams, and other working memory data related to processing. After a time span $\Delta t$, an event  data stream $E$ is stored in the subject's memory. Different modeling layers will have different memory and natural timescales.
 
 
 A Bayesian formulation for Presence is now provided. Consider the probability of a model $M$ co-existing with  a set of events $E$:  
 \begin{equation} \label{eq:prob}
P(E ,M)= P(\mbox{E}| M)\cdot P(M).  
 \end{equation}
 The reality model $M$ is, e.g.,  a model for the self, body, the environment and agency in the environment (the `universe').   
 What this equation says is that {\em the plausibility of a model $M$}  together with  a designed set of events $E$ (the sensorial input plus effector output  stream, to be precise)
 is proportional to the conditional probability of the events given a reality model $M$ (the evidence) times the probability of the designed reality model (the prior, which derives from older data). Here we can formally define two concepts associated to Bayesian terminology, the evidence and the prior:
 the {\em consistency} of new data with the model or {\em evidence} $\sim P(\mbox{Events}| M)$; the {\em possibility} in relation with previous models or  {\em prior} $\sim P(M)$.


Let  $M$ be alternative model of reality intended to be conveyed by a designed VR system. This may be an model targeting  some or all levels of processing. E.g., it may include physical aspects, agency, and target different modeling levels in the hierarchy. The goal of a good Presence system is to convince the user of the `reality' of the experience thorough the joint generation of a set of events (inputs, outputs) $E$.  I note the importance of designed {\em interaction}---both inputs to and outputs from the cognitive system conform the set of events and are important. 

As we will see, there  is no fundamental difference between PI and Psi from the point of view of this mathematical formulation---they both measure a probability. However,  PI will be a function of the  evidence for the desired model low-level aspects as well as its prior probability---which will be rather peaked and not very plastic.  

I define next the formal version Place Illusion associated to an intended model of reality using the Bayesian inference formalism. I will use Greek letters to denote PI ($\Pi$) and Psi ($\Psi$) for short and to remind us of the quantitative and formal shift in their meaning. 
\begin{defi}
 The $M$-Place Illusion  of a model $M$ associated to a set of events $E$ (as generated by interaction of the subject with a virtual reality system with underlying model $M$), $\Pi_M$,  is defined to be 
\begin{equation}
\Pi_M \equiv P(M^l, E) = P(E|M^l)\, P(M^l),
\end{equation}
where $M^l$ refers to the reality model low level (e.g., perceptual) aspects. 
\end{defi}
Since the subject is responsible for a subset of $E$, the model of reality must include him or her as well (e.g., include a  body representation). For hardwired low level models, the prior will dominate.    In order to produce a convincing Place or Body Illusion, the generated events must be reasonably consistent with the intended reality, but this reality must not depart much from the usual one. It is probably very difficult to teach the nervous system to ignore latency problems in vision and touch, or vision-vestibular inconsistencies, for example. It may be possible to believe we have a long nose, but harder to feel we have three eyes. The value of $\Pi_M$ would correspond to the question {\em ``Is the intended low level model real?"}, but this question will probably  not be asked by the subject if he or she are not aware of what is happening or of the intended model.

Similarly, the model-associated Plausibility $\Psi_M$ is proportional to $\Pi_M$,  to the evidence for higher order models,  and to their prior: 
\begin{defi}
The $M$-Plausibility of a model $M$ given a set of events $E$, $\Psi_M$,  is defined to be
\begin{eqnarray}
\Psi_{M} &\equiv&  P(M,  E)  =   P(M^l,E)\cdot P(M^h, E) \nonumber \\
& =& \Pi_M  \cdot  P(E|M^h) \cdot  P(M^h) 
\end{eqnarray}
where we have assumed independence of low and high order models.
\end{defi}
 Presumably, hardwired models are not very affected much by data, but higher level models are plastic and adaptive. 
In the prior $P(M^h)$,  `past-experience' modeling is accounted that will disfavor potential models that depart greatly from those already adopted.  It is for this reason that even in a high realism scenario (digital or not) the subject will not readily accept fully a new reality.  For example, if we suddenly started seeing  flying  cows, it would take us some time to accept this as part of reality.   

The value of $\Psi_M$  corresponds to the question {\em ``Is the intended  model real?"}, but again this question may not be asked by a subject  unaware of what is happening in the VR experience or of the intended model.    
Also note that there is no intrinsic difference between past and present evidence. Past data produces evidence for models, which then become priors for subsequent estimation in a process called sequential estimation \cite{Tipping:2004aa}.

What happens if the $M$ model is poorly designed by the VR engineer, or there are, e.g., timing issues in the system's displays? Then, the intended illusion $M$ will not take place. This could be due to a bad evidence and/or prior (i.e., a low $\Psi_M$ score), or to the existence of a higher plausibility score with an alternative model, $M'$. 
That is, even if the score for $M$ is not high, the subject may still feel that the experience is real. Perhaps he or she will favor  a different reality model than the intended one. In order to account for this  let us define now the generic Plausibility (i.e., not model-tied) of a set of events $E$.
\begin{defi}
The Plausibility of a set of events $E$, $\Psi=\Psi(E)$, is defined to be
\begin{equation}
\Psi = \sum_{M}  P(M, E) =P(E) 
\end{equation}
where the sum is over all models $M$ the subject can construct.
\end{defi}
This would correspond to the question {\em ``Is this really happening?"}.  Perhaps no single model can account well for the data, but there are many potential models out there. For the purpose of model building, it may be sufficient to retain $M_{max}\equiv \max_{M}  P(M, E)$---
{\em ``What is the most plausible model I can construct that fits this experience"}? However, instead of keeping the most probable solution, our brains may work with the $P(M,E)$  distribution function, in a purer Bayesian spirit.  We postulated this occurrence when there is a disagreement between low and high level models (in the rubber-arm experiment). Model independent PI could be defined in a similar way.

  We can write the ideal MDL analog of Equation~\ref{eq:prob} in terms of program length as
  \begin{equation}
  l(M,E)=l(E|M)+l(M)
  \end{equation}
  and we would conjecture that 
\begin{equation}
\Psi =  \sum _{M} 2^{-l(M,E)} \approx 2^{-K_{M,E}}
\end{equation}
where $K_{M,E}$ refers to the total algorithmic length describing the data, which can be decomposed into a sub-program $M$ and  error. If this result were to apply  it would mean that the universal prior is actually used by the cognitive system in question, but this is something to address experimentally. As was argued above, this may well be the case  because simplicity is a practical strategy for cognitive systems, even if the universe is not simple.

 We have stated that low level modeling systems have little memory allocation, so they can only detect simple patterns. It could  be for this reason that PI is an independent phenomenon from Psi. That is, a high $\Psi(E)$ necessitates a high $\Pi(E)$, but not vice-versa.  This hierarchical scheme discussed above is imposed most likely by evolution, since presumably the low level pattern detection mechanisms arose first, in simpler organisms. But this is another experimental question: to what extent can $\Psi$ impact $\Pi$? This possibility is not included in    the simple formalism above, which could be easily extended.
 
We illustrate some of the above ideas with another thought experiment.   Let ARE (`Almost Real') be a real---as opposed to virtual---environment with underlying model $M_W$. The environment is designed to fool the subject into believing they have entered a time machine and travelled in time and space to the Far West, and more specifically into a saloon scene. The subject enters the  room dressed like a cowboy and discovers a  space laid out in full detail, with real people in elaborate   costumes.  The environment  provides a realistic (non-mediated) input/output stream to the subject with a high-level model of being in the Far West (in time and space).  This will yield a  high $\Pi_{M_W}$ (very high place illusion) as there is no illusion in the physical sense---the scenario is real and all the `sensorimotor contingencies' will be met perfectly. Therefore $\Pi_{M_W}= P(E|M^l_W) P(M^l_W) =1 $.  But the subject will not readily believe that the intended situation is real: the Model-Plausibility of $M_W$, $\Psi_{M_W} = \Pi_{M_W} \cdot  P(E|M_W^h) \cdot  P(M_W^h)$ will be greatly  penalized by older data (i.e., the prior) despite the high $\Pi_{M_W}$ value. It will improve as the days or months go by, of course, if the experiment is somehow allowed to continue. The subject, upon entering the room will correctly maintain that ``I am in a  real room with real people in disguise, I am not really in the Far West in the 1800's".  The Plausibility of the events $\Psi(E)$ ( ``Is this happening?") will be high, because there is a well known model  that fits the data and has a reasonable prior (``Somebody is trying to fool me"), albeit not the one intended by the illusion designer.

We close this section with a note on diseases of Presence. Mental disorders  such as schizophrenia, which is marked by dysfunctional  impairments in the perception or expression of reality,   could perhaps be studied as modeling pathologies, given the prevalence of cognitive deficits associated to this disease \cite{Lewis:2004aa}. 




\section{Other applications and future work}
I  provide next an overview of the relevance of simplicity paradigms to other areas, all of which with connections to neuroscience research.

\subsection{Robotics}
The theory of reality construction finds practical applications in robotics. Indeed, if we understand someday how natural brains model the world, we can apply the knowledge to the constructions of robots. And vice-versa, as by trying to build artificial cognitive systems we will probably develop new tools to study the brain. We highlight   some work with connections to the ideas presented here.

The use of self-modeling is an active area of research in robotics. In \cite{Bongard:2006ab} the cognitive agents were robots programmed to  update models of their own bodies (and, extrapolating, their environment) from past 
experience and in this way improve locomotive performance. In particular, this  robot's
 learning cycle included  the design of specially directed experiments to select among 
competing models. This was achieved through Òactuation-sensationÓ interaction with 
the environment. As discussed above, this parallels the scientific method, where 
collected data from directed observations suggest models which are then tested via new experiments. 

In \cite{Philipona:2003aa,Philipona:2004aa} a mathematical model based on the search from simplicity is provided in the context of robotics. The ideas proposed stem from work in \cite{ORegan:2001aa}, where it is argued, much in line with the present work, that the basis of experience arises from the construction of laws relating motor action and sensorial reaction, or more technically, the laws of sensorimotor dependencies. The idea is that such models (which we identify here directly with the concept of `reality') are constructed by exploration and measurement. The point made is that  both motor and sensorial data streams live in very high dimensional spaces. Simplicity must be sought, if nothing else to make control manageable. The authors explain that the existence of invariances in the relation
\begin{equation}
s=\varphi(B(C),u),
\end{equation}
where $u$ is the state of the environment, $C$ is an output motor command which configures the body position $B$, and $s$ is the sensorial input stream---reveals the structure of an underlying group of transformations. To avoid complications the motor command uniquely identifies the body position without reference to past motions. The authors use the language of differential geometry based on the fact that the searched-for models must be invariant with respect to sensorimotor coordinate transformations. Different types of invariances can be explored. For example, invariance with respect to simultaneous motor and environment changes can be used to deduce the dimensionality of `external space' (we use quotes to highlight again  that space is itself a model). We note that the notion of space is directly tied to the type of sensors available. E.g., some `real' dimensions of space may not be accessible at all to the cognitive system in question.  The dimensions of  the invariant space may be deduced by the dimensionality of the sensorial subspaces spanned by motor or environment changes if the sensorial apparatus is blind to them---i.e., if there are invariances.  This work underscores the practical importance of model-building from sensorial information, interaction and searching for simplicity.

\subsection{Fundamental physics}
In physics, information is taking an ever more fundamental role, and since the days in which the physicist John Wheeler posited ``It from bit?" it has become clear that information and the bit could become the modern equivalents of the greek atom and the 20th century quantum. An interesting and modern attempt at reducing physics to information theory is provided in \cite{Frieden:2004aa}, where it is argued that the concept of Fisher information is the key to unification of classical and quantum physics, and a fundamental  physical concept. More recently, a principle called Information Causality which states that {\em  the transmission of m classical bits can cause an information gain of at most m bits} is being studied as a candidate precursor of quantum theory \cite{Pawlowski:2009aa}, thus giving information and information conservation a fundamental status in physics.   In \cite{Ruffini:2007aa}  I discuss in more detail the importance of simplicity and of a brain-centric, information-based approach to the foundations of physics---but see also \cite{Whitworth:2007aa} for an interesting attempt to frame physics in the context of computation. 


\subsection{Biology}

\begin{figure*}[t!]

\vspace{-2.5cm}

\hspace{-.7cm} 
\includegraphics[width=15cm]{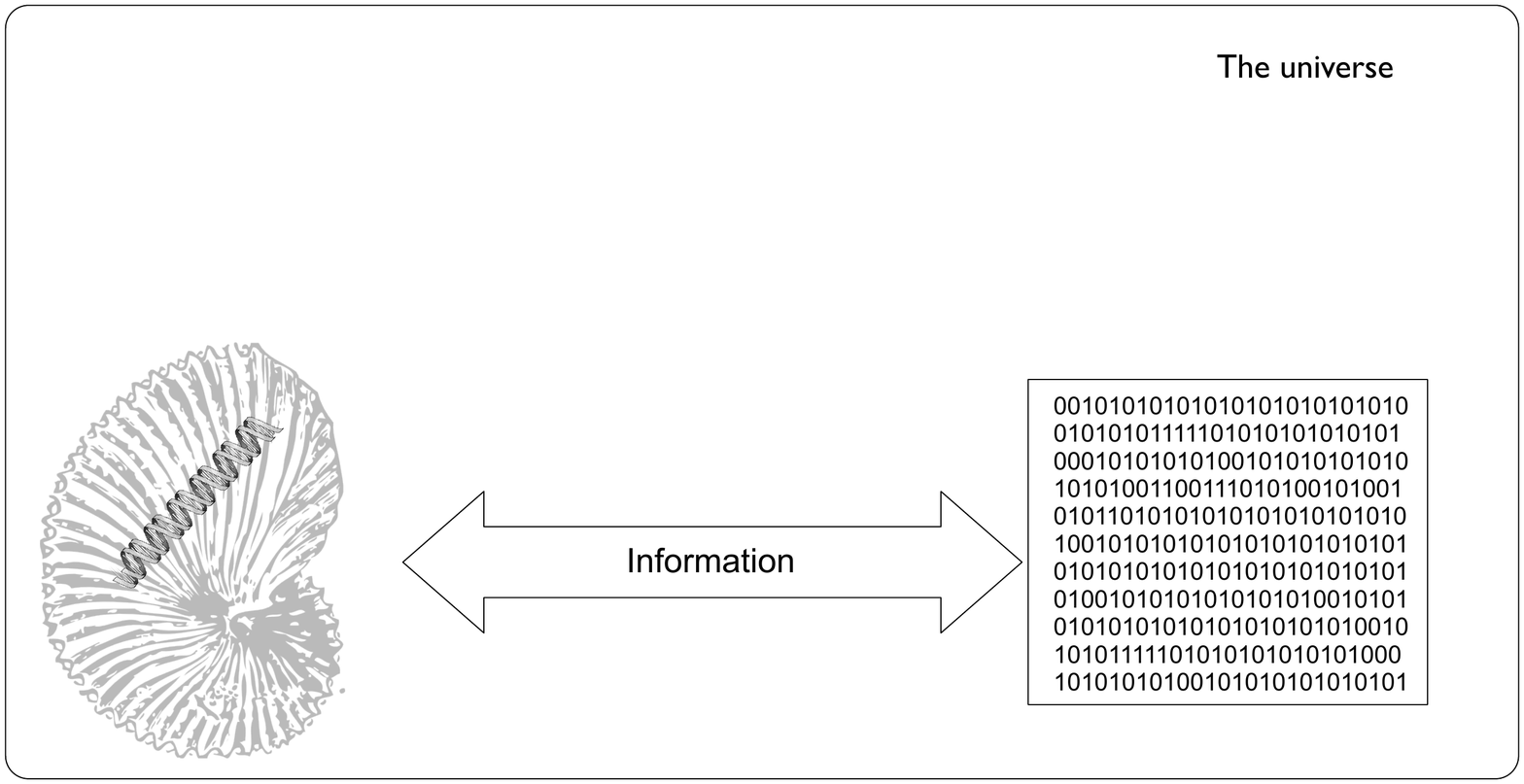}
\vspace{-1.cm}

\caption{An organism creates the model of reality through  information exchange (in and out) with the `outside''. In this case we show the full universe sub-divided into an organism interacting with the rest of the universe by bi-directional information exchange, which physically occurs through  the organism's information membrane. The organism's DNA encodes the model of reality, a model which evolves over long time scales. \label{life} }
\end{figure*}
In the modeling hierarchy, if brains are modeling systems in short time scales, DNA models the environment over long time scales, and these models are manifested (`run') in phenotypes. Organisms can be conceptualized in information-theoretic terms, and evolution as a form of computation. Organisms interact with the environment, and they aim to survive. We can state then that organisms encode models of their environment in their bodies, and ultimately in their DNA. Therefore, DNA can be thought of as a compressed form of the elements in the environment the organism will have to cope with. In the evolutionary long time scales, DNA is the model that encodes those aspects of reality (the environment) needed for survival, which we may call homeostasis with some abuse of language. Figure~\ref{life} provides an illustration of this conceptualization.
Evolution and natural selection thus enter the discourse naturally. If we can define coherently the concept of agents or inference machines at the organism level, it can be argued that those that best compress the information bath will be better prepared to survive. Compression and modeling are  closely tied to prediction and hence survival. 

We can think of evolution as another step in the hierarchical ladder  of computation, but one with a long time scale. E.g., the structure of our bodies reflect a model encoded in our genes that has evolved (as is being  computed) over eons. In this fashion we can extend the idea of cognition from the brain (acting at short time scales) to the organism level (acting on long  time scales through DNA and with evolution).  The discussion above, focusing on the brain interacting with information, can thus be naturally extended to the organism level.

In relation to biology and the origins of the brain, we can state that life and, in particular, nervous systems evolved to find methods to simplify and encapsulate the apparent complexity of the universe, the context  being the usual one: permanence of the fittest (natural selection). Thus, biological evolution is Nature's algorithm for  compression and modeling, a theme to be explored in research and in practical applications.  

In mode of summary, I would like to propose an information-based definition of life spanning all time scales: 
 {\em a living being, or entity or agent,  can be defined to be a replicating program that successfully encodes and runs a (partial) model of reality, thus increasing its chances of survival as a replicating program.}
Evolution is to be thought here of as a form of computation, the means to obtain perduring entities. The results in \cite{Wolpert:2008aa} seem to indicate that there is place for only one strong inference machine in the universe, which could be thought of as the cusp of evolution. Modeling and computation apply, in this definition, at short and long time scales, from DNA to brains.

Evolution is only needed if `perdurance' is under stress. If an organism finds the solution to
perdure without biological evolution (or, at least, natural bio-evolution), aging and mortality are no longer a must.   The stream of reproduction will stop, but the pattern will remain. 

Experimentally, we can try to find evidence for hierarchical modeling and simplicity at the physiological level. The drive for modeling and simplicity through evolution and natural selection, it has been argued, generates hierarchical modeling systems, which at the organic level we may call control systems.  The recent observations of scale-free and multiscale phenomena in physiological signals may be taken as evidence for these ideas, because behind scale free or mutiscale phenomena there are hierarchical generating systems. A natural parallel to the MMN discussion above would be to study simple modeling (control) physiological systems. Simple organisms can provide a very good starting point for such research. Recently, it has been shown that bacteria are also capable of modeling and predicting environmental change \cite{Mitchell:2009aa}, by encoding them in their gene regulatory networks, thus showing that environmental anticipation (modeling) is an adaptive trait even at this biological level.

\subsection{Mathematics and Machine Learning}
The relevance of simplicity to cognitive systems may explain the power of mathematics. Mathematics can be defined as the science of pattern and deductive structure  \cite[p. 108]{Davis:1981aa}. Mathematics  provides the tools for meta-compression: mathematics is the discipline per excellence that draws conclusions logically implied by a set of axioms. Where mathematics is successful  in compressing, a modeling  landscape  is reduced to a finite set of axioms together with equations for `logico-dynamics'. Today we know that any such system will leave gaps and that completeness and consistency are in general incompatible (G\"odel, Turing \cite{Chaitin:1995aa})---how the axioms are chosen is very important.  

Further work is needed to understand the role of algorithmic complexity (KC) in machine learning. Although it has been proven that it is in general  impossible to compute the KC of a string due to the halting problem (or, what amounts to be the same, Godel incompleteness), there may still be something to be proven regarding the performance less demanding tasks. Can we say anything about KC if a limit is imposed on computation time?

A very exciting new trend is emerging in machine learning, the so-called {\em automatization of science} \cite{Schmidt:2009aa, King:2009aa}.  In this groundbreaking work, we find machine learning and simplicity principles coming together for the search for models, but a theory for model building is not provided. Much work remains to be done to optimize the search for models relying on simplicity Bayesian criteria. A natural question in this endeavor is how to choose models that fit the data: how do we trade-off error and model complexity (the work parsimony is normally employed in this context)? With compression as the guiding theme the answer is immediately provided by KC or MDL, which fix this tradeoff and inherently define noise---as discussed at length above.

\subsection{Education}
During development, the most important skill to acquire is arguably the capacity to model or compress information. This is especially true in the 21st century, given the agile access we now have to plentiful data thanks to information technologies. But how do we actually construct high-level abstract models? 
Human beings derive all sorts of models of reality from their interaction with their environment. When a child asks ``why?" he or she is asking for a derivation of a fact from previous knowledge, or perhaps for a new axiom for their model building. This is a fundamental cognitive activity, yet it is poorly understood.  How does the process of modeling  actually take place? Given the fact that we have today a new generation of powerful technologies to control (in the laboratory) the interaction of brains with especially designed environments we can ask how can we use them to explore the transfer of  interaction experiences to more formal knowledge representations. 

Can we conceive means for  cross-hierarchical, cross-modal transfer of modeling skills, e.g., from music to mathematics? Music is already an interesting case: an enjoyable musical piece is a fulfilling exercise in non-trivial model building (which could explain why different people like different types of music---different Bayesian priors).
What is the relation between low level (e.g., reflex) models to higher knowledge? For example, \cite{Smetacek:2004aa}  discusses the relationship of the propioceptive system with our physical modelling capabilities, and hypothesizes that the neural correlates of physics and mathematics did not evolve {\em de novo}, but are rooted in our `subconscious' body sense---proprioception.  Bodily manipulations such as juggling, suggest a well synchronized physical interaction as if the juggler were a physics expert. The juggler uses `embodied knowledge' to interact with the environment. Is this transfer transferred to higher cognitive levels somehow?

\section{Conclusions}

In this paper, following earlier work described in \cite{Ruffini:2007aa},  a neuro-centric, subjective approach to cognition based on information theory has been proposed, with the underlying idea that information is the most fundamental physical and cognitive concept. The discussion is intended to apply to any cognitive system, simple or complex, natural or artificial (e.g., robots).

I have  argued that evolution and natural selection lead to compressing or modeling systems, including auto-modeling. The reason is that the construction of models from data is advantageous for survival.  Modeling is equivalent to compression or the search for simplicity. In this sense, reality, the construction of models from information, is equivalent to simplicity.

I have then reviewed the concept of simplicity and provided mathematical descriptions for this concept.  The search for simplicity, modeling, the search for symmetries, conserved quantities and compression have all been shown to be closely related. Simplicity can be described by Kolmogorov or algorithmic complexity, but also from other angles, such as or the  principle of indifference, Bayes' theory or  minimum description length. A subjective interpretation of probability theory is necessary for it to describe how we   represent knowledge and make decisions.

Although simplicity cannot been proven to be the best strategy for extrapolation and prediction without further assumptions about the universe, it appears to be a practical strategy for compression and unbiased representation of knowledge, allowing for economic storage and manipulation of information from interaction with the environment.  Simplicity is advantageous in deducing, planning, observing, learning, deciding and debugging. It is also advantageous in predicting, if even in a purely practical, resource conscious, way. The NFL theorem implies that, unless we make assumptions about data, we cannot a priori say which algorithm or classifier will perform best. Simplicity is a practical recipe.

From this algorithmic  framework, a hierarchy of compression systems at different levels has been postulated, from those in genes to those in brains,  from low level pattern recognition to complex endeavors including self-models and science. Different approaches can be used to study these sub-systems and their inter-relation. 

I  have demonstrated the use this paradigm with  applications in neuroscience and in current theories of Presence, and shown that  the use of modeling and simplicity as a unifying theme can provide the framework for planning further experimental and theoretical work in these areas. In particular, formal definitions for Plausibility and Place Illusion have been provided using statistical inference and algorithmic information theory concepts. 
I have also proposed that the MMN paradigm, which is used aurally for the study of low level pattern detection in brains, could be expanded to other modalities (somatosensory, visual, etc.) using VR, as well as in the complexity scale to explore model building.

Finally, the  idea of simplicity  as a common unifying thread has been emphasized,  which suggests that a joint inter-disciplinary study approach involving branches of science such as mathematics, machine learning, information theory, neuroscience, physics, computer science, robotics and Presence could be mutually advantageous.





\section*{Acknowledgements}
This work has greatly benefitted  from discussions with many people. Special thanks to Carles Grau, Ed Rietman,  Walter Van de Velde, Mel Slater, David Wolpert, David L. Dowe, Gregory Chaitin and Miriam Reiner, Julian Barbour for their ideas, inspiration and useful discussions. This work partly supported by the PEACH Coordination Action (33909) of the Future and Emerging Technologies (FET) programme within the Sixth Framework Programme for Research of the European Commission.

\bibliographystyle{elsarticle-num}  
\bibliography{kolmogorov}  

\end{document}